\documentclass[twocolumn,amsmath,amssymb]{revtex4}

\usepackage{graphicx}
\usepackage{hyperref}
\begin{document}

\title{A Simple, Direct Finite Differencing of the Einstein Equations}

\date{\today}

\author{Travis M. Garrett}
\email{garrett@phys.lsu.edu}
\affiliation{Louisiana State University}

\begin{abstract}
We investigate a simple variation of the Generalized Harmonic method for
evolving the Einstein equations. 
A flat space wave equation for metric perturbations is separated from the 
Ricci tensor, with the rest of the Ricci tensor becoming a source 
for these wave equations.  
We demonstrate that this splitting method allows for the accurate simulation 
of compact objects, with gravitational field strengths less than or equal to 
those of neutron stars.  
This method could thus provide a
straightforward path for general relativistic effects 
to be added to astrophysics simulations,
such as in core collapse, accretion disks, and extreme mass ratio systems.

\end{abstract}

\maketitle

\newcommand{\be}{\begin{equation}}
\newcommand{\ee}{\end{equation}}

%%%%%%%%%%%%%%%%%%%%%%%%%%%%%%%%%%%%%%%%%%%%%
%  INTRO
%%%%%%%%%%%%%%%%%%%%%%%%%%%%%%%%%%%%%%%%%%%%%
\section{Introduction}

The wave equation is one of the best known 
and most studied 
partial differential equations, and it is 
solved by a variety of numerical methods 
in computer simulations involving radiative problems.
Wave equations 
also appear as the leading order terms within the 
Einstein equations of General Relativity (GR), 
when selecting coordinates 
that follow the harmonic 
coordinate condition $\Box x^a = 0$.
The Generalized Harmonic (GH) formulation of GR 
expands on the harmonic coordinate condition 
by setting the d'Alembertian of the coordinates 
to be a general function: $\Box x^a = H^a$.
This preserves the Einstein equations in a 
wave equation form, but now allows 
for any coordinate system to be used.
The GH formulation of GR was the first to enable the simulation 
of the inspiral, merger and ringdown of 
two black holes \cite{pre05} (and was quickly followed 
by groups using the BSSN formalism \cite{clmz05,bcckm05}
-- for a recent comparison of the two methods see \cite{hannam09}).

We report here on a simple rewriting of the 
GH formulation that further exploits the 
wave equation nature of the Einstein equations.
Flat space wave equations for metric 
perturbations are split off from the Ricci tensor, 
and the rest of the Ricci tensor is 
converted into sources for these 
wave equations.
The complexities of the Einstein field equations are thus packaged 
within these new source terms, which we find to 
be well behaved when the gravitational field strengths
are on the order of those for a neutron star or weaker.
Note that similar methods for splitting off flat 
space wave equations have existed for a long time, 
dating back to de Donder: \cite{dedonder21,dedonder27}
(and see also \cite{fock59,thorne80}).

To determine whether this splitting is 
useful numerically we perform a simple 
simulation of a binary neutron star inspiral and 
merger.
We find that the gravitational fields evolve stably,
that we can extract the expected gravitational waves, 
and we verify that the Hamiltonian 
constraint violations converge to zero.
This test shows that this splitting method 
provides a fairly simple way for 
general relativistic effects and gravitational wave 
production to be added to astrophysics simulations that 
have traditionally only used Newtonian gravity
or simplified versions of GR.
As an example, some current core collapse codes 
(see e.g. \cite{dimmelmeier08}) use the ADM formulation 
of GR, along with the approximation of a conformally flat spatial metric 
 \cite{wilson96, isenberg78}.
Our splitting method provides a simple alternative way to add 
GR, without making any approximations.
It should also be possible to split off and evolve waves equations 
about a curved background (such as Schwarzschild or Kerr),
which would be useful in simulating accretion disks 
or white dwarfs and neutron stars orbiting a supermassive black hole.
 
%%%%%%%%%%%%%%%%%%%%%%%%%%%%%%%%%%%%%%%%%%%%%
%  Field Equations
%%%%%%%%%%%%%%%%%%%%%%%%%%%%%%%%%%%%%%%%%%%%%
\section{Field Equations}

We will briefly review the 
GH formulation of GR -- see e.g. \cite{pre06, lindblom06} for more details.
We start with the metric $g_{ab}$:
\be
ds^2 = g_{ab} dx^a dx^b .
\ee
The Einstein equations $G_{ab} = 8\pi T_{ab}$ can be written in trace reversed form:
\be
R_{ab} = 4 \pi (2 T_{ab} - g_{ab} T) \label{fe1} ,
\ee
with the Ricci tensor $R_{ab}$ given in terms 
of the connection coefficients $\Gamma^c_{ab}$:
\be
R_{ab} = \Gamma^{c}_{ab,c} - \Gamma^{c}_{cb,a} 
+ \Gamma^{d}_{ab} \Gamma^{c}_{dc} - \Gamma^{d}_{cb} \Gamma^{c}_{da} .
\ee
The lowered-indices connection coefficients are given by derivatives 
of the metric:
\be
\Gamma_{abc} = \frac{1}{2} ( g_{ab,c} + g_{ac,b} - g_{bc,a}) ,
\ee
and the metric can be used to raise the first index or get a contracted form:
\begin{eqnarray}
\Gamma^{a}_{bc} &=& g^{ad} \Gamma_{dbc} ,\\ 
\Gamma^a &=& g^{bc} \Gamma^a_{bc} ,\\
\Gamma_a &=& g^{bc} \Gamma_{abc} .
\end{eqnarray}

The contracted connection coefficients are 
equivalent to minus a wave operator acting on the coordinates:
\be
g^{bc} \nabla_b \nabla_c x^a = \Box x^a  = -\Gamma^a .
\ee
Choquet-Bruhat showed \cite{bruhat52} that if one 
picks harmonic coordinates $\Box x^a = 0$,
then the Ricci tensor can be expanded and 
the Einstein equations (\ref{fe1}) transformed into:
\begin{align}
g^{cd}g_{ab,cd} + 2 g^{cd}{}_{(,a} g_{b)d,c} 
+ 2 \Gamma^d_{cb} \Gamma^c_{da}  \label{fe2} \\ \notag
 = -8 \pi (2T_{ab} - g_{ab} T) .
 \end{align}
 The highest order term in Eq. (\ref{fe2}) is the curved space 
 wave equation $g^{cd}g_{ab,cd}$, so this places the 
 Einstein equations in a manifestly hyperbolic form.

A wider range of options are available however -- one has the 
freedom in General Relativity to pick any set of coordinates.  
As shown by Friedrich \cite{friedrich85} and Garfinkle \cite{garfinkle02}
this freedom can be alternatively expressed by 
picking source functions $H^a (x,t)$ to drive 
the wave equations for the coordinates:
\be
\Box x^a = H^a (x,t) = -\Gamma^a \label{hadef} .
\ee
Using these source functions we can decompose Eq. (\ref{fe1})
in the standard GH formulation:
\begin{align}
g^{cd}g_{ab,cd} + 2 g^{cd}{}_{(,a} g_{b)d,c} 
+ 2H_{(a,b)} - 2 H_c \Gamma^c_{ab} \label{fe3} \\ \notag
 + 2 \Gamma^d_{cb} \Gamma^c_{da} = -8 \pi (2T_{ab} - g_{ab} T) .
 \end{align}

We will now modify this equation in order to explicitly 
separate off the wave equations, 
and convert the rest of the Ricci tensor into sources.
We will proceed generally at first by considering metric perturbations
$f_{ab}$ away from a general background metric $\bar{g}_{ab}$, 
(such as a Minkowski or Schwarzschild spacetime):
\be
g_{ab} = \bar{g}_{ab} + f_{ab} .
\ee
We then refer to the raised indices perturbations 
by $h^{ab}$:
\be
g^{ab} = \bar{g}^{ab} + h^{ab}
\ee
(note that the $h^{ab}$ perturbations
are determined by the background metric $\bar{g}_{ab}$ 
and the lowered indices perturbations $f_{ab}$).

One can then split the second order piece of Eq. (\ref{fe3}) into:
\be
g^{cd}g_{ab,cd} = (\bar{g}^{cd} + h^{cd})(\bar{g}_{ab,cd} + f_{ab,cd}) .
\ee
We want to separate out a scalar wave equation for each of the metric 
perturbations $f_{ab}$ on the background geometry $\bar{g}_{ab}$:
\be
\bar{g}^{cd} \bar{\nabla}_c \bar{\nabla}_d f_{ab} = \bar{g}^{cd} f_{ab,cd} 
- f_{ab,e} \bar{\Gamma}^e .
\ee
We thus need to subtract $f_{ab,e}\bar{\Gamma}^e$ from both sides 
of Eq. (\ref{fe3}) in order to get $\bar{\Box} f_{ab}$ on the LHS.
We collect the rest of the terms on the LHS of Eq. (\ref{fe3}) 
into a source $S_{ab}$ which we will move to the RHS:
\begin{align}
S_{ab} = -\bar{g}^{cd}\bar{g}_{ab,cd} - h^{cd}(\bar{g}_{ab,cd} + f_{ab,cd}) \label{seq1} \\ \notag
 -2 g^{cd}{}_{(,a} g_{b)d,c} - 2 H_{(a,b)} + 2 H_c \Gamma^c_{ab} \\ \notag
- 2 \Gamma^d_{cb} \Gamma^c_{da} - f_{ab,e} \bar{\Gamma}^e . \\ \notag
\end{align}
This allows us to rewrite the standard GH equations 
(\ref{fe3}) in the simple form:
\be
\bar{\Box} f_{ab} = S_{ab} -8 \pi (2T_{ab} - g_{ab} T) \label{fe4} .
\ee

In Eqs. (\ref{seq1}) and (\ref{fe4}) we give a splitting 
for metric perturbations on a general background 
geometry $\bar{g}_{ab}$.  Here we specialize to the 
case examined in this paper: we set the background 
to be flat and use Cartesian coordinates 
$\eta_{ab} = diag(-1,1,1,1)$:
\be
g_{ab} = \eta_{ab} + f_{ab} \label{hc1} .
\ee
In this case the coordinate source functions 
$H^a$ are zero and the $S_{ab}$ source terms simplify to:
\begin{align}
S_{ab} = - h^{cd} f_{ab,cd} -2 g^{cd}{}_{(,a} g_{b)d,c} - 2 \Gamma^d_{cb} \Gamma^c_{da} \label{seq2}
\end{align} 
(where all of the metric derivatives now stem from the $f_{ab}$ functions).

%%%%%%%%%%%%%%%%%%%%%%%%%%%%%%%%%%%%%%%%%%%%%
%  Numerical Simulation
%%%%%%%%%%%%%%%%%%%%%%%%%%%%%%%%%%%%%%%%%%%%%
\section{Numerical Simulation}

\subsection{Matter EOM}

In order to test our splitting (\ref{seq1}-\ref{fe4})
of the Einstein equations and see if they allow 
for stable and accurate numerical evolutions, 
we build a simplistic model of a binary neutron 
star inspiral and merger.
We construct and evolve a stress energy tensor 
$T_{ab}$ by fiat during the simulation, which  
sources the gravitational fields $f_{ab}$ (which 
in turn give rise to the corrective sources $S_{ab}$).
Note that we are not using $T^{ab}{}_{;b} = 0$ to generate realistic 
equations of motion for the matter, as the 
main goal for this (single processor) code is to test 
the response of the gravitational fields 
$f_{ab}$ to the rapid accelerations of dense 
matter sources.
Later versions could use quasiequilibrium binary 
neutron star initial data as in \cite{baumgarte97,koji99}, 
or fully evolve the matter equations of motion as in 
\cite{anderson07}.

The neutron stars are constructed as rigid polytropes 
with a compactness in the range of: $M/R \sim 0.1-0.3$.
We choose to drive these density profiles 
on a quasi-circular inspiral path as found by Peters and Mathews 
\cite{peters63,peters64}, which captures the leading order 
radiation reaction effects.
The inspiral is parameterized by:
\be
a(t) = a_0 \left(1-\frac{t}{t_{decay}} \right)^{1/4} ,
\ee
where $a(t)$ is the semimajor axis as a function 
of time, $a_0$ is the initial semimajor axis (between 
the centers of masses of the bodies), and the 
decay time is given by:
\be
t_{decay} = \frac{5}{64} \frac{a_0^4}{M^3} \label{tdecay} ,
\ee
where $M$ is the total mass of the 
(equal mass) neutron stars.  
The instantaneous coordinate velocities of the stars are simply 
given by the Keplerian velocities of a binary 
with the same mass and separation.
Note that eventually the separation $a(t)$
will decrease enough that the stars begin to overlap: here the 
stress energy tensor is determined by a simple superposition 
of the individual stars densities, with the separation 
quickly falling to zero and the velocities also 
modulated to zero.  This is, of course, not a particularly 
realistic model of a binary neutron star merger, 
the point is just to demonstrate that the gravitational 
fields $f_{ab}$ still evolve stably in this 
strong field and highly dynamical setting.

Having determined the bulk motion of the stars, 
we need the distributions of their fluid variables in 
order to build the stress energy tensor.
We choose the standard perfect fluid form:
\be
T^{ab} = (\rho(1+\varepsilon) + p)u^a u^b + p g^{ab} ,
\ee
with rest mass density $\rho$, specific internal 
energy $\varepsilon$, pressure $p$, and four 
velocities $u^a$.  The internal energy and 
pressure are given in terms of the density 
through a polytropic equation of state:
\begin{eqnarray}
p &=& \kappa \rho^{\Gamma}  , \\
\varepsilon &=& \frac{\kappa}{\Gamma - 1} \rho^{\Gamma - 1} ,
\end{eqnarray}
where we choose $\Gamma = 2$, and a value for $\kappa$ 
such that the internal energy density is about 
$5 \%$ of $\rho$ in the center of the star.  

The density 
$\rho$ is in turn derived from a conserved 
``baryonic" density $\rho^{\ast}$ (see e.g. \cite{tegp}):
\be
\rho^{\ast} = \rho (-g)^{1/2} u^0 .
\ee
This density has a number of convenient properties, 
including a flat space conservation law:
\be
\partial_t \rho^{\ast} = \partial_i (v^i \rho^{\ast}) ,
\ee
which ensures that the total integrated baryonic mass:
\be
M^{\ast} = \int \rho^{\ast} d^3 x ,
\ee
is constant throughout the evolution.  We thus choose 
initial radial profiles for $\rho^{\ast}$ for each star and 
hold these constant during the evolution.

%%%%%%%%%%%%%%%%%%%%%%%%%%%%%%%%%%%%%%%%%
%  Field Evolution
%%%%%%%%%%%%%%%%%%%%%%%%%%%%%%%%%%%%%%%%%
\subsection{Field Evolution}

Having determined the form of the matter stress 
energy tensor, we now concentrate on how to evolve 
the gravitational fields $f_{ab}$ in response via 
(\ref{fe4}) (which has been simplified by Eqs. (\ref{hc1}-\ref{seq2})).  
We proceed in a slightly nonstandard 
way: given previous successes we choose to finite 
difference the scalar wave equation $\Box f_{ab}$
using implicit methods, while we use 
the standard explicit methods to construct the 
source $S_{ab}$ (note that this is a particular choice, but 
any other stable scheme to evolve wave equations should work as well).

First consider the implicit finite differencing 
of $\Box f_{ab}$.  We use operator 
splitting to divide the 3+1 wave equation into three 
1+1 wave equations
(see e.g. \cite{recipes}).  For conciseness 
replace $f_{ab}$ with $\psi$ and the entire source
$S_{ab} - 8 \pi (2T_{ab} - g_{ab}T)$ with $\tau$.
Adopting an operator splitting strategy we divide:
\be
-\partial^2_t \psi + \partial^2_x \psi
+ \partial^2_y \psi + \partial^2_z \psi = \tau
\ee
into a set of 1+1 problems:
\begin{align}
-\partial^2_t \psi + \partial^2_x \psi = 1/3 \tau , \label{i11} \\ 
-\partial^2_t \psi + \partial^2_y \psi = 1/3 \tau , \\
-\partial^2_t \psi + \partial^2_z \psi = 1/3 \tau ,
\end{align}
which are sequentially finite-differenced implicitly.
We use a variation of the Crank-Nicholson method, modified 
for second time derivatives, to evolve these -- 
for instance Eq. (\ref{i11}) becomes:
\begin{eqnarray}
\frac{1}{\Delta t^2} (\psi^{N+1}_i - 2 \psi^N_i + \psi^{N-1}_i)  &=& \label{fd1} \\ \notag
 c_{N+1} \frac{1}{\Delta x^2}
 (\psi^{N+1}_{i+1} &- 2 \psi^{N+1}_i& + \psi^{N+1}_{i-1}) \\ \notag
+ c_{N-1} \frac{1}{\Delta x^2}
(\psi^{N-1}_{i+1} &- 2 \psi^{N-1}_i& + \psi^{N-1}_{i-1}) 
- \frac{1}{3} \tau^N_i ,
\end{eqnarray}
where the field is discretized at time step $N$ and 
spatial mesh location $i$.
We have also included generalized coefficients $c_{N+1}$ and 
$c_{N-1}$ (with $c_{N+1}+c_{N-1}=1$) 
so that we can transition from Crank-Nicholson 
$c_{N+1} = c_{N-1} = 1/2$ to a fully implicit method 
$c_{N+1} = 1, c_{N-1} = 0$, or somewhere in between.
More implicit splitting allows us to dissipate 
high frequency noise if needed.
Boundary conditions are added to Eq. (\ref{fd1})
which is then solved by inverting the resulting tridiagonal matrix.
We note that in general implicit methods allow one 
to take as large a time step $\Delta t$ as desired and still maintain 
stability, however for reasons of accuracy we take 
time steps that are about the same size 
an explicit scheme takes in order to satisfy  
the Courant stability condition.

This operator-splitting implicit method is combined 
with a Fixed Mesh Refinement (FMR) grid 
structure so that a high resolution mesh is available 
near the origin to resolve the stars, while also extending
through successive lower resolution meshes 
out into the wave zone.
Sommerfeld outgoing wave boundary conditions 
are applied to the largest and coarsest mesh, which is  
updated first.  Successively finer meshes are then 
updated, with their boundary conditions found via interpolation 
from the next coarsest mesh.  After all the meshes are 
updated one time step the coarser mesh points 
are replaced with the finer mesh's field values 
wherever there is overlap (all meshes are updated 
each time step -- no subcycling in time is used).

This leaves the evaluation of the metric source $S_{ab}$ given 
by Eq. (\ref{seq2}).  The spatial derivatives 
are computed in the standard second order 
explicit way, but evaluating the time derivatives is more subtle 
as the updated field values $f^{N+1}_{ab}$ will not be known 
until $\Box f_{ab}$ is calculated.  We choose to solve this iteratively:
for the first pass the first and second time derivatives contained 
in (\ref{seq2}) are evaluated using the current and two 
previous time steps $f^{N}_{ab}$, $f^{N-1}_{ab}$, 
$f^{N-2}_{ab}$ -- this allows for a preliminary 
value of $\hat{f}^{N+1}_{ab}$ to be found.
The second iterative pass then uses the preliminary 
$\hat{f}^{N+1}_{ab}$
to calculate centered time derivatives in (\ref{seq2})
which are used to complete the time step.

Finally initial data is needed -- we choose very simple 
initial data: $f_{ab}(t=0) = 0$.  There is thus a large 
amount of noise initially as the fields respond to the 
matter and then settle down to their correct physical 
solutions, with the transients propagating off the grid
within a fraction of an orbit.  
For values of neutron star 
compactness much larger than $M/R \sim 0.1$ it is 
useful to also add a large amount of dissipation initially, 
and then linearly decrease the dissipation to a small or zero value over a 
fraction of an orbit.  
Otherwise the large amplitudes of the initial transient spikes 
can lead to infinities in the $S_{ab}$ sources (for example if  $|f_{00}| \ge 1$), 
thus crashing the code.
Using physical initial data could also solve this, 
but it is encouraging that the code still rapidly converges to the 
correct solution when given initial data which completely 
ignores the self gravity of the stars 
(as has been seen elsewhere, see e.g. \cite{bishop05}).

%%%%%%%%%%%%%%%%%%%%%%%%%%%%%%%%%%%%%%%%%
%  Results
%%%%%%%%%%%%%%%%%%%%%%%%%%%%%%%%%%%%%%%%%
\subsection{Results}

\begin{figure}
\includegraphics[width=3.5in]{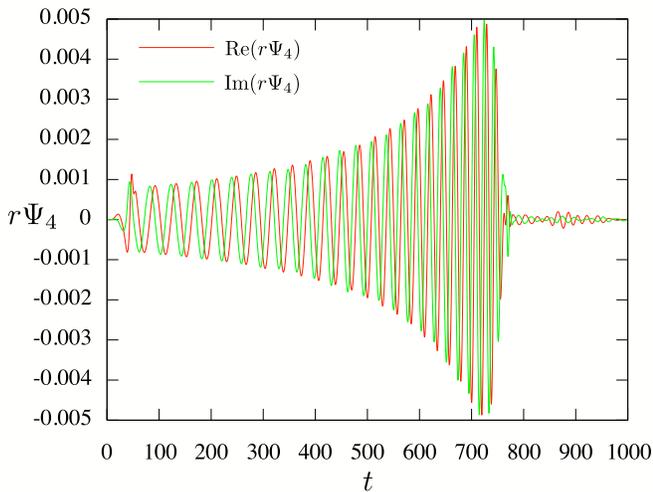}
\caption{The $l=m=2$ mode of $\Psi_4$ as a function of time,
demonstrating the standard chirp waveform.}
\end{figure}

We find that our method of splitting off flat space 
wave equations for metric perturbations and turning the 
rest of the Ricci tensor into sources provides an 
effective method for simulating
systems with neutron star strength or weaker gravitational fields.
The results for an example evolution are given: we pick 
two stars, each with baryonic mass $M^{\ast} = 0.15$ 
and coordinate radius $R=1$, and with an initial coordinate 
separation of 4 (which gives a value of $t_{decay} \sim 751$ via Eq. (\ref{tdecay})).
The finest mesh spacing is $\Delta x = 0.2$, giving about 10 grid 
points across the diameter of the stars.  

Initially the Hamiltonian constraint:
\be
{}^3R - K_{ij}K^{ij} + K^2 - 16 \pi \rho_H = 0 \label{hamcon}
\ee
is highly violated as all the terms except
for the Hamiltonian density $\rho_H$ are zero.
As mentioned, after a short while the initial transients propagate off the grid
and the Ricci scalar ${}^3R$ grows to 
counterbalance $\rho_H$ (with the 
$K_{ij}K^{ij}$ and $K^2$ terms giving small corrections).
The constraint (\ref{hamcon}) is then satisfied to within several percent, 
which is improved further with increasing resolution.

In figure (1) the real and imaginary 
parts of the $l=2$, $m=2$ mode of the gravitational waveform $\Psi_4$
are shown
(as measured at a radial distance of $50$ or $\sim 167 M^{\ast}$).
The initial noise propagates through, 
and the waveform then settles down approximately to the standard chirp profile.
After the merger the waveform dies away quickly, 
due to the abrupt cessation of motion for the artificially driven sources
(there is a small amount of residual noise visible, which 
converges to zero with increasing resolution).

The simple binary inspirals we have evolved show that 
our direct method in equation (\ref{fe4}) 
successfully solves the Einstein equations, 
at least in the case of neutron star type systems with a flat Minkowski 
background (\ref{hc1})-(\ref{seq2}).
This method thus provides a simple way for 
general relativistic effects to be added to 
astrophysical simulations.

\begin{acknowledgments}
We would like to thank Luis Lehner, Carlos Palenzuela, and Charles Evans for 
useful discussions.  This work was supported in part by 
NSF grants PHY-0803629, PHY-0653375, PHY-0653369, and CCF-0833193.
Computations were done at LSU and LONI, and on Teragrid through grants 
MCA09X003, and MCA02N014.
\end{acknowledgments}

\bibliographystyle{unsrt}
\bibliography{genh2}

\begin{thebibliography}{10}

\bibitem{pre05}
F.~Pretorius.
\newblock Evolution of binary black hole spacetimes.
\newblock {\em Phys.Rev.Lett.}, 95:121101, (2005), gr-qc/0507014.

\bibitem{clmz05}
M.~Campanelli, C.~O. Lousto, P.~Marronetti, and Y.~Zlochower.
\newblock Accurate evolutions of orbiting black-hole binaries without excision.
\newblock {\em Phys. Rev. Lett.}, 96:111101, (2006), gr-qc/0511048.

\bibitem{bcckm05}
J.~G. Baker, J.~Centrella, D.~I. Choi, M.~Koppitz, and J.~van Meter.
\newblock Gravitational wave extraction from an inspiraling configuration of
  merging black holes.
\newblock {\em Phys. Rev. Lett.}, 96:111102, (2006), gr-qc/0511103.

\bibitem{hannam09}
H.~Hannam, S.~Husa, J.~G. Baker, M.~Boyle, B.~Bruegmann, T.~Chu, N.~Dorband,
  F.~Herrmann, I.~Hinder, B.~J. Kelly, L.~E. Kidder, P.~Laguna, K.~D. Matthews,
  J.~R. van Meter, H.~P. Pfeiffer, D.~Pollney, C.~Reisswig, M.~A. Scheel, and
  D.~Shoemaker.
\newblock The samurai project: verifying the consistency of black-hole-binary
  waveforms for gravitational-wave detection.
\newblock {\em arXiv:0901.2437v2 [gr-qc]}, (2009).

\bibitem{dedonder21}
T.~DeDonder.
\newblock La gravifique einsteinienne.
\newblock {\em Gunthier-Villars, Paris}, (1921).

\bibitem{dedonder27}
T.~DeDonder.
\newblock The mathematical theory of relativity.
\newblock {\em Massachusetts Institute of Technology, Cambridge, MA}, (1927).

\bibitem{fock59}
V.~Fock.
\newblock {\em The Theory of Space Time and Gravitation}.
\newblock Pergamon Press, New York, (1959).

\bibitem{thorne80}
K.~S. Thorne.
\newblock Multipole expansion of gravitational radiation.
\newblock {\em Rev. Mod. Phys.}, 52:341--392, (1980).

\bibitem{dimmelmeier08}
H.~Dimmelmeier, C.~D. Ott, A.~Marek, and H.-T. Janka.
\newblock The gravitational wave burst signal from core collapse of rotating
  stars.
\newblock {\em arXiv:0806.4953v2 [astro-ph]}, (2008).

\bibitem{isenberg78}
J.~A. Isenberg.
\newblock Waveless approximation theories of gravity.
\newblock {\em arXiv:gr-qc/0702113v1}, (1978).

\bibitem{wilson96}
J.~R. Wilson, G.~J. Mathews, and P.~Marronetti.
\newblock Relativistic numerical method for close neutron star binaries.
\newblock {\em Phys. Rev. D}, 54:1317--1331, (1996), arXiv:gr-qc/9601017.

\bibitem{lindblom06}
L.~Lindblom, M.~A. Scheel, L.~E. Kidder, R.~Owen, and O.~Rinne.
\newblock A new generalized harmonic evolution system.
\newblock {\em Class. Quant. Grav.}, 23:S447--S462, (2006), gr-qc/0512093.

\bibitem{pre06}
F.~Pretorius.
\newblock Simulation of binary black hole spacetimes with a harmonic evolution
  scheme.
\newblock {\em Class. Quant. Grav.}, 23:S529--S552, (2006), gr-qc/0602115.

\bibitem{bruhat52}
Y.~Four\`es-Bruhat.
\newblock Th\'eor\`eme d'existence pour certains syst\`emes d'\'equations aux
  d\'eriv\'ees partielles non lin\'eaires.
\newblock {\em Acta Math}, 88:141--225, (1952).

\bibitem{friedrich85}
H.~Friedrich.
\newblock On the hyperbolicity of einstein's and other gauge field equations.
\newblock {\em Commun. Math. Phys.}, 100:525--543, (1985).

\bibitem{garfinkle02}
D.~Garfinkle.
\newblock Harmonic coordinate method for simulating generic singularities.
\newblock {\em Phys. Rev. D.}, 65:044029, (2002), gr-qc/0110013.

\bibitem{koji99}
K.~Uryu and Y.~Eriguchi.
\newblock New numerical method for constructing quasiequilibrium sequences of
  irrotational binary neutron stars in general relativity.
\newblock {\em Phys. Rev. D}, 61:124023, (2000), gr-qc/9908059v2.

\bibitem{baumgarte97}
T.~W. Baumgarte, G.~B. Cook, M.~A. Scheel, S.~L. Shapiro, and S.A. Teukolsky.
\newblock General relativistic models of binary neutron stars in
  quasiequilibrium.
\newblock {\em Phys. Rev. D}, 57:7299--7311, (1998), gr-qc/9709026.

\bibitem{anderson07}
M.~Anderson, E.~W. Hirschmann, L.~Lehner, S.~L. Liebling, P.~M. Motl,
  D.~Neilsen, C.~Palenzuela, and Tohline~J. E.
\newblock Simulating binary neutron stars: dynamics and gravitational waves.
\newblock {\em Phys. Rev. D}, 77:024006, (2008) arXiv:0708.2720.

\bibitem{peters63}
P.~C. Peters and J.~Mathews.
\newblock Gravitational radiation from point masses in a keplerian orbit.
\newblock {\em Phys. Rev. 131,435}, (1963).

\bibitem{peters64}
P.~C. Peters.
\newblock Gravitational radiation and the motion of two point masses.
\newblock {\em Phys. Rev. 136,1224}, (1964).

\bibitem{tegp}
C.~M. Will.
\newblock {\em Theory and Experiment in Gravitational Physics Revised Edition}.
\newblock Cambridge University Press, (1993).

\bibitem{recipes}
W.~H. Press, S.~A. Teukolsky, Vetterling~W. T., and B.~P. Flannery.
\newblock {\em Numerical Recipes in C 2nd Ed.}
\newblock Cambridge University Press, (1992).

\bibitem{bishop05}
N.~T. Bishop, R.~Gomez, L.~Lehner, M.~Maharaj, and J.~Winicour.
\newblock On characteristic initial data for a star orbiting a black hole.
\newblock {\em Phys. Rev. D}, 72:024002, (2005) arXiv:gr-qc/0412080.

\end{thebibliography}

\end{document}